\documentclass[numberedappendix,twocolumn,twocolappendix]{openjournal}

\usepackage[table]{xcolor}

\usepackage{latexsym}
\usepackage{graphicx}
\usepackage{amssymb}
\usepackage{longtable}
\usepackage{epsf}
\usepackage{amsmath}
\usepackage{graphicx}
\usepackage{textcomp}
\usepackage{gensymb}

\usepackage{hyperref}
\hypersetup{colorlinks=true,linkcolor=blue,citecolor=blue,filecolor=blue,urlcolor=blue}
\usepackage{cleveref}
\usepackage{bm}

\usepackage{natbib}
\usepackage{orcidlink}

\usepackage{xspace}

\newcommand{\msun}{{\rm M}_{\odot}}

\newcommand{\beq}{\begin{eqnarray}}
\newcommand{\eeq}{\end{eqnarray}}
\newcommand{\ben}{\begin{itemize}}
\newcommand{\een}{\end{itemize}}

\newcommand{\Mhost}{M_{\rm host}}
\newcommand{\mhost}{m_{\rm host}}

\newcommand{\mobs}{m_{\rm obs}}

\newcommand{\tobs}{t_{\rm obs}}
\newcommand{\Mpeak}{M_{\rm peak}}

\newcommand{\msub}{m_{\rm sub}}
\newcommand{\mhalo}{m_{\rm halo}}

\newcommand{\thetaMAH}{\theta_{\rm MAH}}
\newcommand{\thetacosmo}{\theta_{\rm cosmo}}

\newcommand{\Mhalo}{M_{\rm halo}}
\newcommand{\nhalo}{n_{\rm halo}}
\newcommand{\nhalocuml}{N_{\rm halo}}

\newcommand{\Msub}{M_{\rm sub}}
\newcommand{\nsub}{n_{\rm sub}}
\newcommand{\nsubcuml}{N_{\rm sub}}

\newcommand{\colossus}{{\tt Colossus}\xspace}
\newcommand{\halox}{{\tt Halox}\xspace}
\newcommand{\diffhalos}{{\tt Diffhalos}\xspace}
\newcommand{\diffmah}{{\tt Diffmah}\xspace}

\newcommand{\diffmahnet}{{\tt DiffmahNet}\xspace}
\newcommand{\jaxcosmo}{{\tt jax-cosmo}\xspace}
\newcommand{\jax}{{\tt jax}\xspace}
\newcommand{\DstarPop}{{DiffstarPop}\xspace}

\definecolor{hpurple}{HTML}{7E16DF}

\definecolor{horange}{HTML}{FFA500}

\definecolor{hred}{HTML}{d62728}

\begin{document}

\title{Diffhalos: A Generative Model of Cosmological Lightcones of Dark Matter Halos\vspace{-1.5cm}}
\shorttitle{Diffhalos: A generative model of halo lightcones}

\author{Georgios Zacharegkas$^{1,\star}$\orcidlink{0000-0002-2890-6758}}
\author{Andrew P. Hearin$^1$\orcidlink{0000-0003-2219-6852}}
\author{Alan Pearl$^1$}
\author{Matthew R. Becker$^1$\orcidlink{0000-0001-7774-2246}}
\author{Florian Kéruzoré$^1$}
\author{Sara Ortega-Martinez$^{2}$}

\affiliation{$^1$HEP Division, Argonne National Laboratory, 9700 South Cass Avenue, Lemont, IL 60439, USA}
\affiliation{$^2$Donostia International Physics Center, Manuel Lardizabal Ibilbidea, 4, 20018 Donostia, Gipuzkoa, Spain}
\thanks{$^{\star}$E-mail: gzacharegkas@anl.gov}

\shortauthors{Zacharegkas et al.}

%%%%%%%%%%%%%%%%%%%%%%%%%%%%%%%%%%%%%%%%%%%%%%%%%%
\begin{abstract}
We present a generative model of cosmological lightcones of dark matter halos, \diffhalos. In our model, we draw Monte Carlo samples of the halo mass function in a lightcone with a JAX-based implementation of the halo model, \halox, and we generate samples of subhalos by drawing from a model for the conditional subhalo mass function. We generate mass assembly histories (MAHs) using a normalizing flow trained on merger trees in cosmological N-body simulations. We show that \diffhalos can generate samples of halos, subhalos, and their MAHs with a statistical distribution that accurately approximates populations in simulated lightcones. As an example application, we use \diffhalos to calculate gradients of the halo and subhalo mass functions with respect to cosmological parameters. We conclude with a discussion of ongoing work using \diffhalos together with models of the galaxy--halo connection to make theoretical predictions for cosmological populations of galaxies, and to generate mock galaxy catalogs.
\end{abstract}
%%%%%%%%%%%%%%%%%%%%%%%%%%%%%%%%%%%%%%%%%%%%%%%%%%

\maketitle

\vspace{1cm}

\twocolumngrid

\section{Introduction}
\label{sec:intro}

In cold dark matter (CDM) cosmologies, most of the dark matter is organized in halos, which form the fundamental unit of structure formation. Galaxies form and evolve at the centers of dark matter halos \citep{white_rees_1978,blumenthal_etal_1984}, and thus observable properties of galaxies are closely connected to the properties of the halos that host them \citep[see, e.g.,][]{somerville_dave_2015,Wechsler_Tinker_2018_review}.
Modern galaxy surveys are a powerful probe of large-scale structure cosmology, and present-day and near-future cosmology surveys will measure the properties of billions of galaxies spanning most of cosmic time.

Lightcones of synthetic galaxies are one of the core simulated data products used to support contemporary cosmological analyses of galaxy surveys. Such lightcones can be generated on-the-fly as a gravitational N-body simulation is run \citep{Evrard_etal2001,Fosalba_etal2007,pichon_etal2010_aski,Fosalba_etal2013,Springel_etal2020}, or assembled in post-processing by interpolating between a collection of output snapshots \citep[e.g.,][]{merson_etal2013_lightcone,Hollowed_etal2019,Sgier_etal2020,izquierdo_etal2023,Hadzhiyska_etal2023}, often requiring massively parallel codes and I/O pipelines \citep[e.g.,][]{heitmann_etal21_last_journey,Ramakrishnan_2025}. 

When generating synthetic galaxy catalogs, theoretical predictions can be made with greater richness and physically realistic complexity if the simulated halos include information about halo mass assembly history (MAH). The dark matter halo MAH lies at the foundation of contemporary models of galaxy formation physics: semi-analytic models (SAMs) treat the halo MAH as the driving source of fresh material for star formation \citep[e.g.,][]{bower_durham_sam_2006,somerville_etal_2008,henriques_etal_2015,lagos_shark_2018}; semi-empirical approaches use the halo MAH as a fundamental input to scaling relations of the galaxy--halo connection  \citep{becker_smad_2015,moster_emerge1, behroozi_etal19}; even empirical models such as abundance matching require MAH-derived properties in order to successfully fit target summary statistics of large-scale structure \citep{reddick_etal13,hearin_watson_2013,chaves_montero_etal16_sham_eagle,contreras_etal21_shame}. The assembly of a halo is characterized by its {\em merger tree}, which describes how its mass is built up through the accretion of ``progenitors" that merge into the main halo over time. Halo merger trees are by now a standard data product of modern N-body simulations, and a wide range of codes are available to construct merger trees from simulated cosmological volumes \citep[e.g.,][]{behroozi_etal13_consistent_trees,rodriguez_gomez_etal15_sublink,poole_etal17_gbptrees,mansfield_etal24_symfind,Kong_etal2025_sidm}.

A widely used alternative to simulated MAHs are models of halo merger trees. The prevailing framework for such models is the (extended) Press-Schechter formalism \citep[][EPS, hereafter]{press_schechter_1974, bower_1991, lacey_cole_93,zentner_eps_2007}; synthetic merger trees based on EPS have long been used in the pipelines of SAMs \citep[e.g.,][]{somerville_etal_2008}, and continue to provide a useful basis for predictions of the observed galaxy distribution \citep[e.g.,][]{driskell_etal24}. More recently, machine learning techniques have been developed to generate samples of halo MAHs \citep[e.g.,][]{robles_etal22,nguyen_etal24_florah,nguyen_etal25}.

An alternative approach to generating samples of halo MAHs was introduced in \citet{hearin_etal21_diffmah} (hereafter H21), in which the MAH of an individual halo is approximated with the \diffmah fitting function with 4 parameters, $\thetaMAH$. Then, for a dark matter halo with redshift-zero mass $M_0$, a parametric model for the probability distribution $P(\thetaMAH \vert M_0)$ was calibrated against halo MAHs in N-body simulations.

The \diffmah model is an example of a growing literature on differentiable programming in scientific computing, in which modern libraries of automatic differentiation are used to implement theoretical predictions and simulations, offering the benefits of both gradient information as well as GPU acceleration. The \diffmah model is implemented in  \jax \citep{jax2018github}, one of the more widely used autodiff libraries in astrophysics and cosmology. For example, the \jaxcosmo \citep{Campagne_etal_jaxcosmo} and LINX \citep{Giovanetti_2025_linx} libraries were built on \jax and enable differentiable calculations of core predictions in cosmology. Differentiable dark matter halo properties and large-scale-structure calculations for halo modeling are provided by the library \halox \citep{Keruzore_halox}. \jax has also recently been used as the basis of models of the baryonic gas content of dark matter halos \citep{pandey_etal25_godmax}, models of stellar population synthesis \citep{Hearin_2023_dsps,martin_navarro_faster_2026}, semi-analytic models of galaxy formation \citep{Pandya_etal2026_sapphire}, and hydrodynamical simulations \citep{horowitz_etal25_diffhydro}.

In this paper, we introduce \diffhalos, a model for generating cosmological lightcones, where each halo and subhalo in the simulated population has a mass assembly history based on \diffmah. We leverage a differentiable implementation of the halo mass function (HMF) to generate a lightcone of host halos; we explore differentiable HMFs  based on \halox, as well as based on a custom emulator optimized for lightcone generation. We introduce a differentiable model for the subhalo mass function for efficiently generating subhalo populations. Finally, we have trained a normalizing flow architecture to approximate the probability distribution of the MAHs of the halos in a lightcone, generalizing the model in H21. We provide publicly available code for generating Monte Carlo realizations of halo lightcones, as well as memory-efficient, Quasi-Monte Carlo generators.

This paper is organized as follows. \S\ref{sec:sims} describes the simulations we use to train and validate our models. In \S\ref{sec:overview}, we give a brief and high-level overview of the \diffhalos framework. We describe the model of the halo mass functions we use in \S\ref{sec:HMF} in detail. Following that, we introduce our host halo lightcone generator in \S\ref{sec:lightcone_generator} and our subhalo lightcone generator in \S\ref{sec:subhalo_generator}. We discuss \diffmahnet, our new model for generating halo mass assembly histories in the lightcones in \S\ref{sec:diffmahnet}. In \S\ref{sec:discussion}, we discuss our results and outline future applications, and we conclude with a brief summary in \S\ref{sec:conclusion}.

%%%%%%%%%%%%%%%%%%%%%%%%%%%%%%%%%%%%%%%%%%%%%%%%%%

\section{Simulations and Merger Trees}
\label{sec:sims}

To develop and validate our models for the mass assembly history (MAH) of halos, we used Small MultiDark Planck (SMDPL), a gravity-only N-body simulation that belongs to the series of MultiDark simulations \citep{riebe_etal13_multidark_database}. As described in \citet{klypin_etal16}, SMDPL was carried out using the GADGET-2 code \citep{springel_2005_gadget2} by evolving $3840^3$ dark-matter particles of mass $m_{\rm p}=9.6\times10^{7}\msun$ in a box of $400\,{\rm Mpc}$ on a side under cosmological parameters closely matching \citet{planck14b}. For the halo MAHs, we used publicly available\footnote{\url{https://www.peterbehroozi.com/data.html}} merger trees that were identified with Rockstar and ConsistentTrees \citep{behroozi_etal13_rockstar, behroozi_etal13_consistent_trees, rodriguez_puebla_etal16}.

\diffmah is designed to capture the {\em cumulative peak} halo mass, $\Mpeak(t):$ the largest mass the halo has ever attained up until time $t.$ By definition, $\Mpeak(t)$ cannot decrease; for example, during a period of time over which the instantaneous halo mass drops, $\Mpeak(t)$ remains constant. Capturing halo mass loss due to flyby events and post-infall subhalo evolution requires additional modeling ingredients beyond our present scope (although see \S\ref{sec:discussion} for further discussion), and so throughout the paper when we refer to $\Mhalo$ or ``halo mass", we are referring to $\Mpeak(t).$ 

In the text and figures throughout the paper, values of mass and distance are quoted using the value of the dimensionless Hubble constant $h$ of the assumed cosmology (i.e., masses are quoted in $\msun$, not in $\msun/h).$

\subsection{Notation}\label{sec:notation}

Throughout the paper, we use the term ``subhalo" to refer to a smaller halo that resides inside the radius of a more massive ``host" halo. When referring specifically to the mass of the larger host halo associated with a subhalo, we use the notation $\Mhost$ to denote the mass of the parent halo. Similarly, when referring specifically to the mass of the smaller subhalo, we use the subscript ``sub" and denote its mass by $\Msub.$ For example, as described in \S\ref{sec:subhalo_generator}, our model for the abundance of subhalos is defined in terms of $\mu \equiv \Msub / \Mhost,$ the mass ratio of a subhalo to its host. Finally, when referring to an object that could either be a host halo or a subhalo, we use the variable $\Mhalo,$ and the general term ``halo mass".

We use lower-case $\mhalo \equiv \log_{10} \Mhalo$ for the base--10 logarithm of halo mass, to help declutter notation in some equations and figures. Similarly, we write $\mhost \equiv \log_{10} \Mhost$ and $\msub \equiv \log_{10} \Msub$ for host and subhalos, respectively.

%%%%%%%%%%%%%%%%%%%%%%%%%%%%%%%%%%%%%%%%%%%%%%%%%%

\section{Diffhalos Overview}
\label{sec:overview}

In this section, we give a high-level overview of \diffhalos. To generate a lightcone of halos, subhalos, and their mass assembly histories, we factor the problem into three separate stages: i) generate a population of host halos in the lightcone volume; ii) populate each halo with subhalos; iii) paint a mass assembly history onto each halo and subhalo. Below we sketch our approach to each of these three steps, referring to subsequent sections for further details.

First, to generate a population of host halos in a lightcone, we assume a parametric form of the halo mass function (HMF), $n_{\rm halo}(\mhalo,z),$ and use a two-dimensional form of inverse transformation sampling to generate a Monte Carlo realization of this analytic HMF. This approach to lightcone generation benefits from a novel parameterization of the HMF in terms of the {\em cumulative} abundance of halos, $\nhalocuml(>\mhalo,z),$ described in detail in \S\ref{sec:HMF}; additional technical details about our halo lightcone generation appear in \S\ref{sec:lightcone_generator}.

Second, we paint a population of subhalos within each host halo in the lightcone. Our model for subhalo generation is based on a parametric form for the conditional subhalo mass function (CSHMF), $\nsub(\mu \vert \mhost)$, where $\mu$ is the ratio of the mass of the subhalo to that of its host, $\mu \equiv \Msub / \Mhost.$ Again we adopt a cumulative form of the CSHMF parameterization that is especially convenient for generating Monte Carlo realizations; further details about generating subhalos appear in \S\ref{sec:subhalo_generator}. At this stage, we have a population of halos and subhalos, each with a redshift, $z$, and a mass at that redshift, $\Mhalo(z)$. 

Finally, we map a mass assembly history (MAH) to each (sub)halo in the lighcone. Our approach is based on the \diffmah model of halo MAH\footnote{In the present work, we use an updated 5-parameter version of the \diffmah model that includes a new parameter $t_{\rm peak}$ controlling arrested growth \citep[see][Appendix A]{Alarcon_2025_diffstarpop}}, in which $\Mhalo(t)$ is approximated with a smooth functional form parameterized by $\thetaMAH$. Under this approximation, the task of generating a MAH reduces to mapping the parameters $\thetaMAH$ onto each (sub)halo. We accomplish this mapping using \diffmahnet, a neural density estimation of $P(\thetaMAH \vert z, \Mhalo(z)),$ described in detail in \S\ref{sec:diffmahnet}.

This three-stage design of \diffhalos enables flexibility in how the library can be used in downstream applications. For example, if an N-body simulation were used to generate a lightcone of host halos, the HMF implementation in \diffhalos could be ignored, the subhalo generator could be used to populate the simulated host halos with subhalos, and then \diffmahnet could bestow a MAH onto each (sub)halo in the resulting catalog. See \S\ref{sec:discussion} for further discussion.

%%%%%%%%%%%%%%%%%%%%%%%%%%%%%%%%%%%%%%%%%%%%%%%%%%

%%%%%%%%%%%%%%%%%%%%%%%%%%%%%%%%%%%%%%%%%%%%%%%%%%

\section{Halo Mass Function}
\label{sec:HMF}

Generating a cosmological lightcone of dark matter halos requires calculating the halo mass function (HMF), which quantifies the abundance of halos as a function of mass and redshift. The HMF varies as a function of cosmological parameters, $\thetacosmo,$ and in this paper we explore two different techniques for achieving differentiability in our HMF calculations: one based on a differentiable implementation of the analytical halo model, and another based on an emulator. We discuss these two techniques in turn below.

\subsection{Halo model calculation}
\label{subsec:halox_hmf}

In the halo model, the $\thetacosmo$-dependence of the HMF is inherited through its calculation in terms of the linear matter power spectrum, $P_{\rm m}(k, z \vert \thetacosmo).$ The power spectrum is convolved with a window function to compute  $\sigma^2(R, z),$ the mass variance inside a Lagrangian radius $R = [3M / (4\pi \Omega_m \rho_{\rm c})]^{1/3}$ enclosing mass $M$, where $\rho_{\rm c}$ is the critical density and $\Omega_m$ is the matter density parameter:
\begin{equation}\label{eq:sigmaR}
    \sigma^2(R, z) = \frac{1}{2\pi^2} \int dk \; k^2 P_{\rm m}(k, z \vert \thetacosmo) \vert \tilde{W} (kR) \vert^2 \; ,
\end{equation}
where $\tilde{W} (kR)$ is the Fourier transform of the real-space window function $W$. The mass function $\nhalo$ is then calculated in terms of the derivative of the mass variance:
\begin{equation}\label{eq:HMFdef}
    \nhalo(\mhalo, z) = f(\sigma) \frac{\Omega_m \rho_{\rm c}}{\Mhalo} \frac{{\rm d} \ln \sigma^{-1}}{{\rm d} \ln \Mhalo} \; .
\end{equation}

\begin{figure}
\centering
\includegraphics[width=\columnwidth]{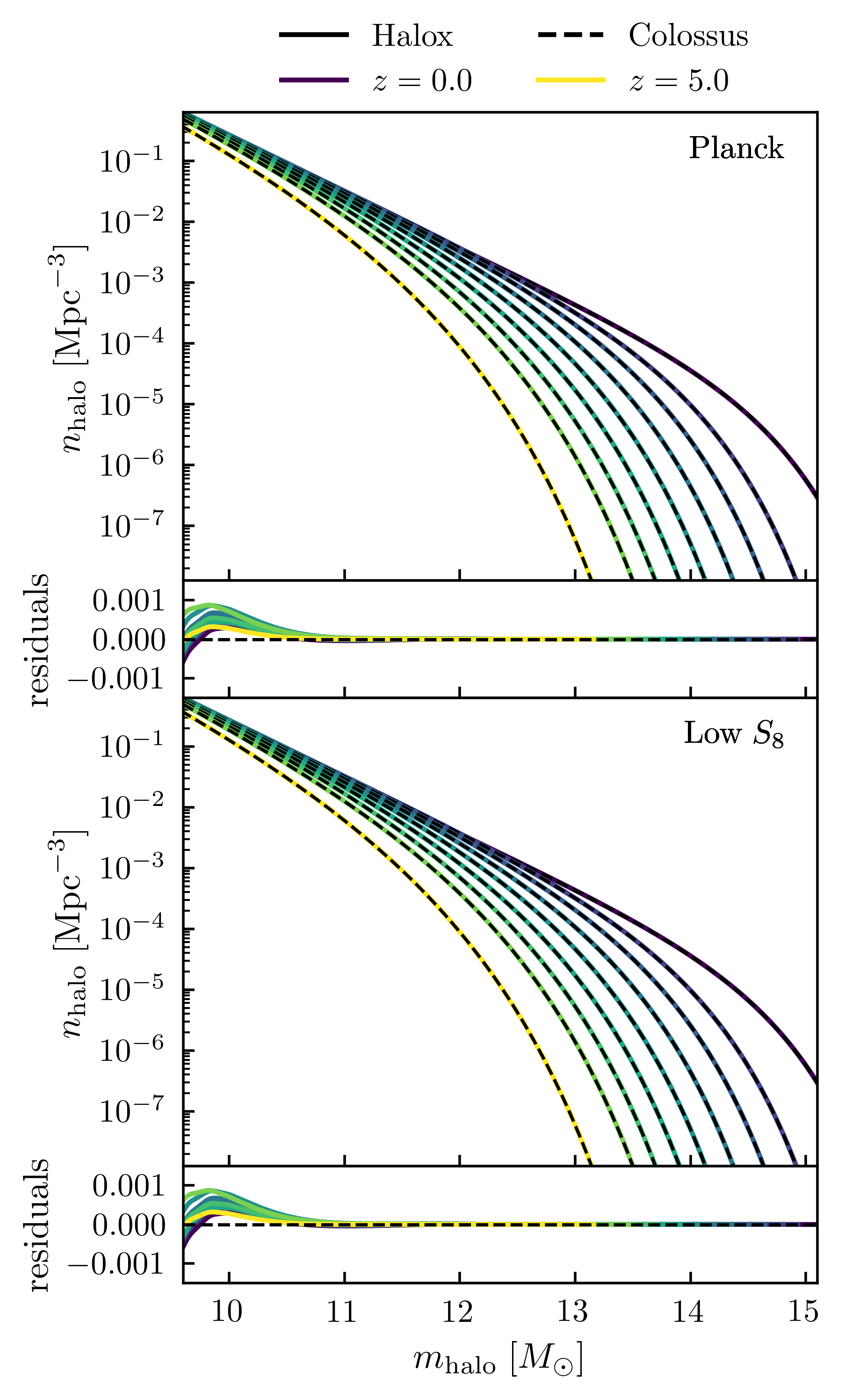}
\caption{
    \label{fig:HMF_model_halox_vs_colossus}
    Comparison between HMF predictions from \halox (solid) and \colossus (dashed) for two different cosmologies: Planck 2018 on the top and a low $S_8$ cosmology on the bottom. Below each panel we also show the fractional differences between the two models. Predictions are compared for redshifts ranging from $z=0$ (purple) to $z=5$ (yellow).}
\end{figure}

The \colossus library\footnote{\url{https://bdiemer.bitbucket.io/colossus/index.html}} provides a publicly available implementation of halo model calculations such as the HMF, halo bias, and a wide range of halo profile models. Calculations of the HMF in \colossus are highly efficient and have a flexible Python interface, but the computations are not implemented in autodiff, and so it is not possible to directly use \colossus in the differentiable generation of halo populations. We instead rely on \halox, a programmatic implementation of the halo model in \jax \citep{Keruzore_halox}. \halox supports cosmology dependence of its predictions through \jaxcosmo \citep{Campagne_etal_jaxcosmo}, a \jax-based library for some of the core calculations in theoretical cosmology. Both \halox and \colossus implement the \citet{tinker_etal08_mass_function} fitting formula for $f(\sigma)$ to predict the mass function, and in Figure~\ref{fig:HMF_model_halox_vs_colossus} we demonstrate percent-level agreement between the two calculations over a wide range of halo mass and redshift, showing comparisons at two different cosmologies: Planck 2018 in the top and a low $S_8\equiv \sigma_8 \sqrt{\Omega_m/0.3}$ cosmology in the bottom panel.

%%%%%%%%%%%%%%%%%%%%%%%%%%%%%%%
%%%%%%%%%%%%%%%%%%%%%%%%%%%%%%%

\begin{figure}
\centering
\includegraphics[width=\columnwidth]{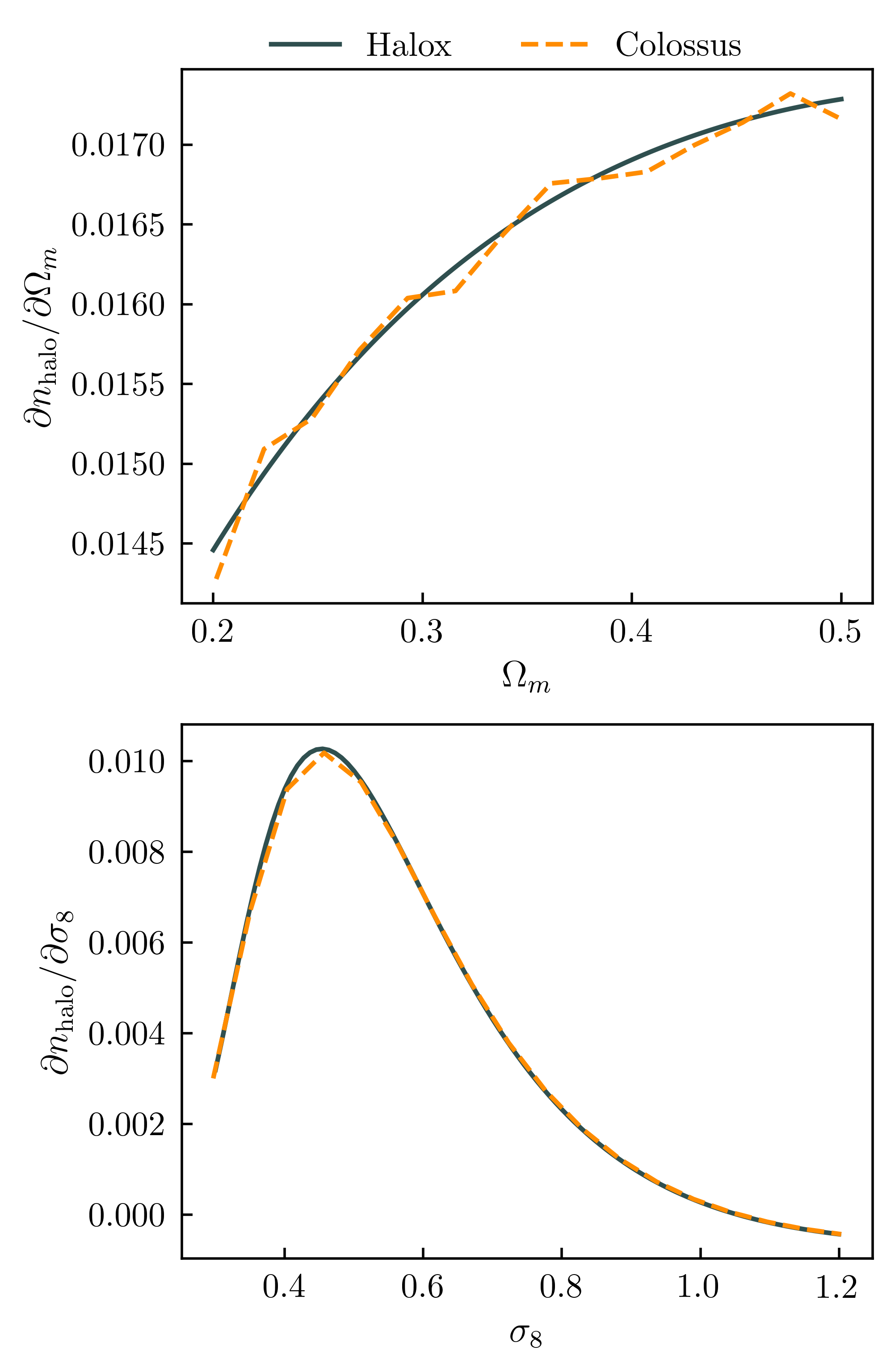}
\caption{
    \label{fig:HMF_grad_halox_vs_colossus}
    Comparison between predictions of HMF gradients with respect to the cosmological parameters $\Omega_m$ (top) and $\sigma_8$ (bottom) using \halox (solid gray) and \colossus (dashed orange).
}
\end{figure}

The \jax-based implementation of \halox enables the use of autodiff to compute gradients of the HMF with respect to cosmological parameters, ${\rm d}\nhalo/{\rm d}\thetacosmo$. In Figure~\ref{fig:HMF_grad_halox_vs_colossus}, we compare the \halox calculations of these gradients to finite-difference estimations based on \colossus; the top panel shows this comparison for the case of $\Omega_m,$ and the bottom panel shows $\sigma_8,$ the amplitude of mass fluctuations on scales of $8 \; h^{-1} {\rm Mpc}$. We refer the reader to \citet{Keruzore_halox} for further discussion of \halox.

%%%%%%%%%%%%%%%%%%%%%%%%%%%%%%%
%%%%%%%%%%%%%%%%%%%%%%%%%%%%%%%

\subsection{Emulator-based calculation}
\label{subsec:ai_hmf}

The halo model calculation outlined in \S\ref{subsec:halox_hmf} is a flexible technique for capturing the cosmology-dependence of the HMF, but using this technique in the \diffhalos pipeline to generate cosmological samples of halos can lead to a heavy memory footprint. As discussed further in \S\ref{sec:lightcone_generator} below, generating a Monte Carlo realization of the HMF with inverse transformation sampling requires self-consistently evaluating both $\nhalo(\mhalo, z)$ and the {\em cumulative} halo mass function, $\nhalocuml(>\mhalo, z).$ Thus in the halo model approach, two numerical integrals are required: one to compute the mass variance (Eq.~\ref{eq:sigmaR}), and one to compute $\nhalocuml(>\mhalo,z)$ from $\nhalo(\mhalo,z).$ We find that vectorizing these integrals to generate halo populations using {\tt jax.vmap} consumes large memory resources even for modestly sized halo populations. It may be possible to trim the memory footprint of this approach to the calculation by carefully tuning the lookup tables used in the mass variance integrations, and/or by alternative algorithms to vectorization such as {\tt jax.lax}, but in this section we explore an alternative approach to the calculation: directly emulating $\nhalocuml(>\mhalo,z),$ and then using {\tt jax.grad} to ensure a self-consistent calculation of $\nhalo(\mhalo,z).$ 

In the emulator-based approach, we first directly parametrize the cumulative halo mass function, $\nhalocuml(>\mhalo, z)$, as a power-law with a slope that smoothly varies with mass according to a \textit{sigmoid function} \textendash a fitting function shape we refer to as \textit{sig-slope}. The differential HMF $\nhalo(\mhalo, z)$ is then computed by applying \texttt{jax.grad} to the parametric fit of the cumulative mass function. Thus, in this approach, the parameters specifying an individual HMF are the parameters of the sig-slope fitting function, $\theta_{\rm HMF}.$ Finally, to capture cosmology dependence, we use a Multilayer Perceptron (MLP) to approximate the mapping between $\thetacosmo$ and $\theta_{\rm HMF}.$

\begin{figure}
\centering
\includegraphics[width=\columnwidth]{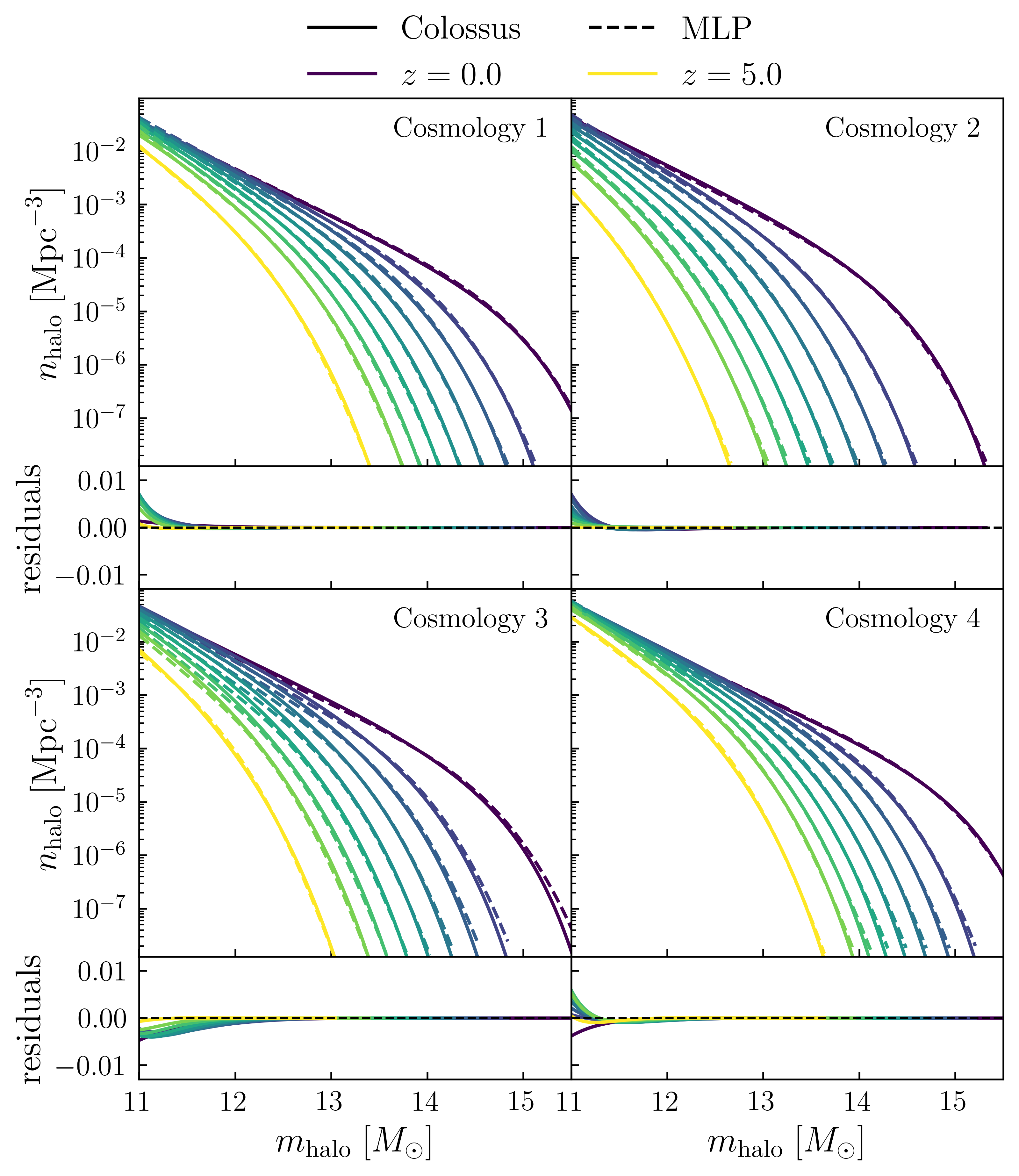}
\caption{\label{fig:HMF_model_mlp_vs_colossus}
    Comparison between HMF predictions from our MLP-based model (dashed) and \colossus (solid), for different cosmologies in each panel. Below each panel we also show the differences between the two models. Predictions are compared for redshifts ranging from $z=0$ (purple) to $z=5$ (yellow).}
\end{figure}

In Figure~\ref{fig:HMF_model_mlp_vs_colossus}, we demonstrate the accuracy of an example calibration of an MLP that we have trained by using \colossus to generate training data for a range of two cosmological parameters, $\Omega_m$ and $\sigma_8$, across the redshift range $0<z<5$, and halo mass range $10^{11}\msun<\Mhalo<10^{15}\msun$. In this example, halo masses are defined as $M_{\rm 200m}$: the mass enclosed within a radius where the average interior density is $200$ times the mean matter density of the Universe. Our example MLP achieves an accuracy of $10\%$ or better at the full range of halo mass and redshift we consider. The \cite{tinker_etal08_mass_function} mass function itself is only accurate across the $10-20\%$ level, and so we have not attempted to achieve a higher-accuracy MLP with further training. The aim of the present paper is merely to introduce the differentiable methodology underlying \diffhalos, not to supply new percent-level emulators, but in \S\ref{sec:discussion} we outline a program for turning \diffhalos into a tool for precision cosmology in future work. 

%%%%%%%%%%%%%%%%%%%%%%%%%%%%%%%
%%%%%%%%%%%%%%%%%%%%%%%%%%%%%%%

%%%%%%%%%%%%%%%%%%%%%%%%%%%%%%%%%%%%%%%%%%%%%%%%%%

\section{Lightcone generator}
\label{sec:lightcone_generator}

In this section, we describe our methodology for generating Monte Carlo (MC) realizations of lightcones of host halos based on the analytical HMFs defined in \S\ref{sec:HMF}.

First, we use inverse transformation sampling to assign a redshift to a Monte Carlo realization of host halos in the lightcone. For a thin shell at redshift $z_{\rm shell}$ spanning $(z_{\rm lo},z_{\rm hi}),$ we can compute $N_{\rm shell}\propto \nhalocuml(>M_{\rm min},z_{\rm shell}) - \nhalocuml(>M_{\rm max}, z_{\rm shell}),$ the total number of halos in the shell with masses in the range $(M_{\rm min}, M_{\rm max}).$ We tabulate a grid $N_{\rm shell}(z_{\rm i}),$ for a grid $\{z_{\rm i}\},$ and sum the grid to calculate $N_{\rm tot},$ the total number of halos generated in the entire lightcone volume. We generate redshifts for the halos via inverse transformation sampling: we use the grid $N_{\rm shell}(z_{\rm i})$ to numerically define the inverse cumulative function $N^{-1}_{\rm tot}(>z);$ we draw $N_{\rm tot}$ uniform randoms $u_{\rm j}$ from the unit interval, and use the inverse cumulative function to calculate the redshift $z_{\rm j}$ associated with each $u_{\rm j}.$

%%%%%%%%%%%%%%%%%%%%%%%%%%%%%%%
\begin{figure}
\centering
\includegraphics[width=\columnwidth]{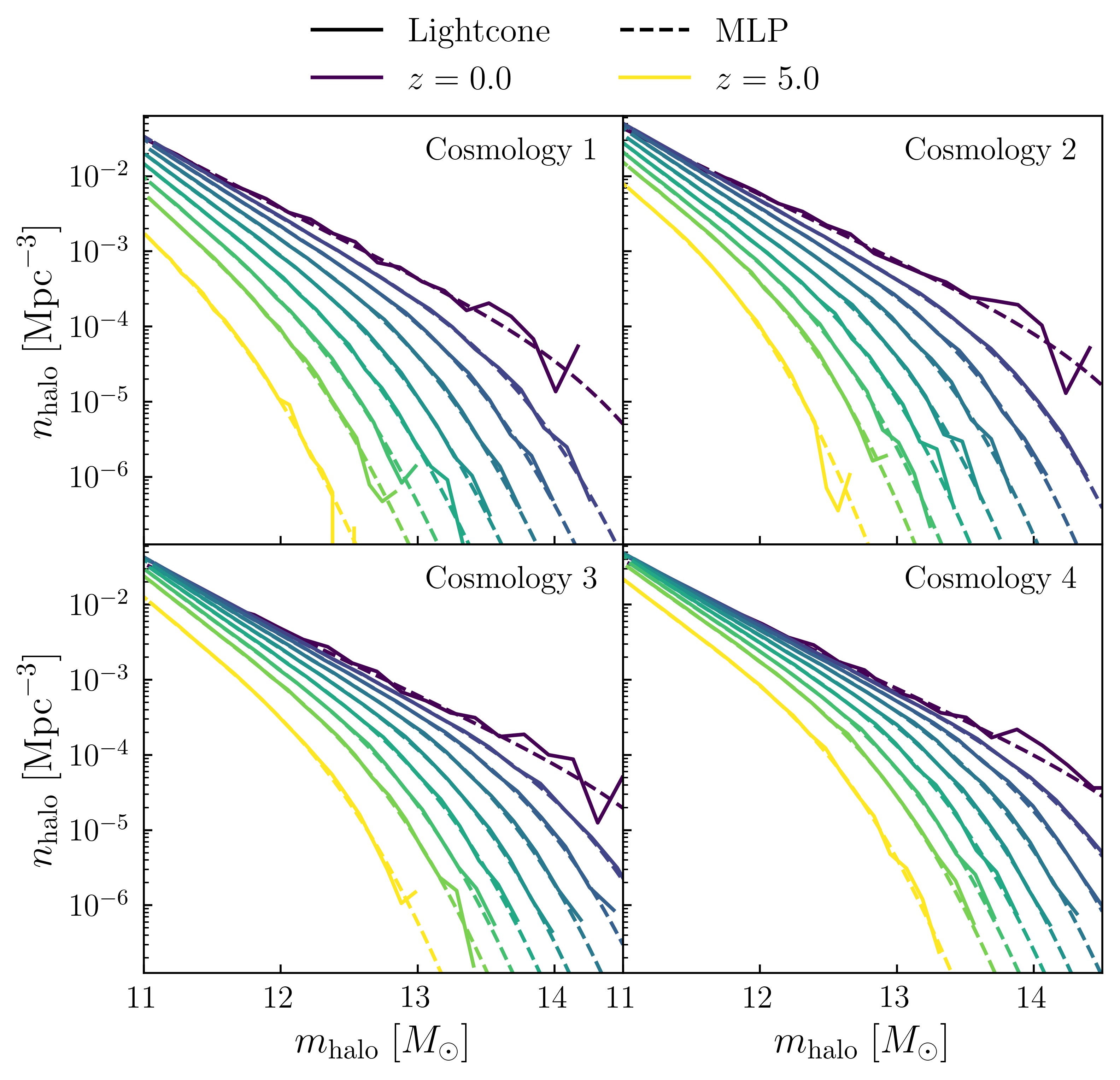}
\caption{\label{fig:HMF_model_vs_MCgen} Comparison between the Monte Carlo realizations (solid) of the HMF and the analytical model predictions (dashed). Different panels correspond to different cosmologies, while different colored lines represent different redshifts ranging from $z=0$ to $z=5$, as indicated in the legend.}
\end{figure}
%%%%%%%%%%%%%%%%%%%%%%%%%%%%%%%

The above application of inverse transformation sampling bestows each of the $N_{\rm tot}$ halos in our MC-generated population with a value of redshift; in a subsequent application of the same technique, we map values of halo mass onto the population. For each individual halo in our sample with redshift $z_{\rm halo},$ we can numerically define the inverse cumulative function $\nhalocuml^{-1}(>\mhalo, z_{\rm halo}),$ which allows us to associate a uniform random $u$ with a halo mass. Using {\tt jax.vmap} is a computationally efficient way to scale this numerical inversion to a large halo population. In Figure~\ref{fig:HMF_model_vs_MCgen}, we show that the statistical distribution of MC-generated halos agrees with the underlying analytic HMF model, over a wide range of halo mass and redshift, and for different cosmological parameters in each panel.

%%%%%%%%%%%%%%%%%%%%%%%%%%%%%%%
\begin{figure}
\centering
\includegraphics[width=\columnwidth]{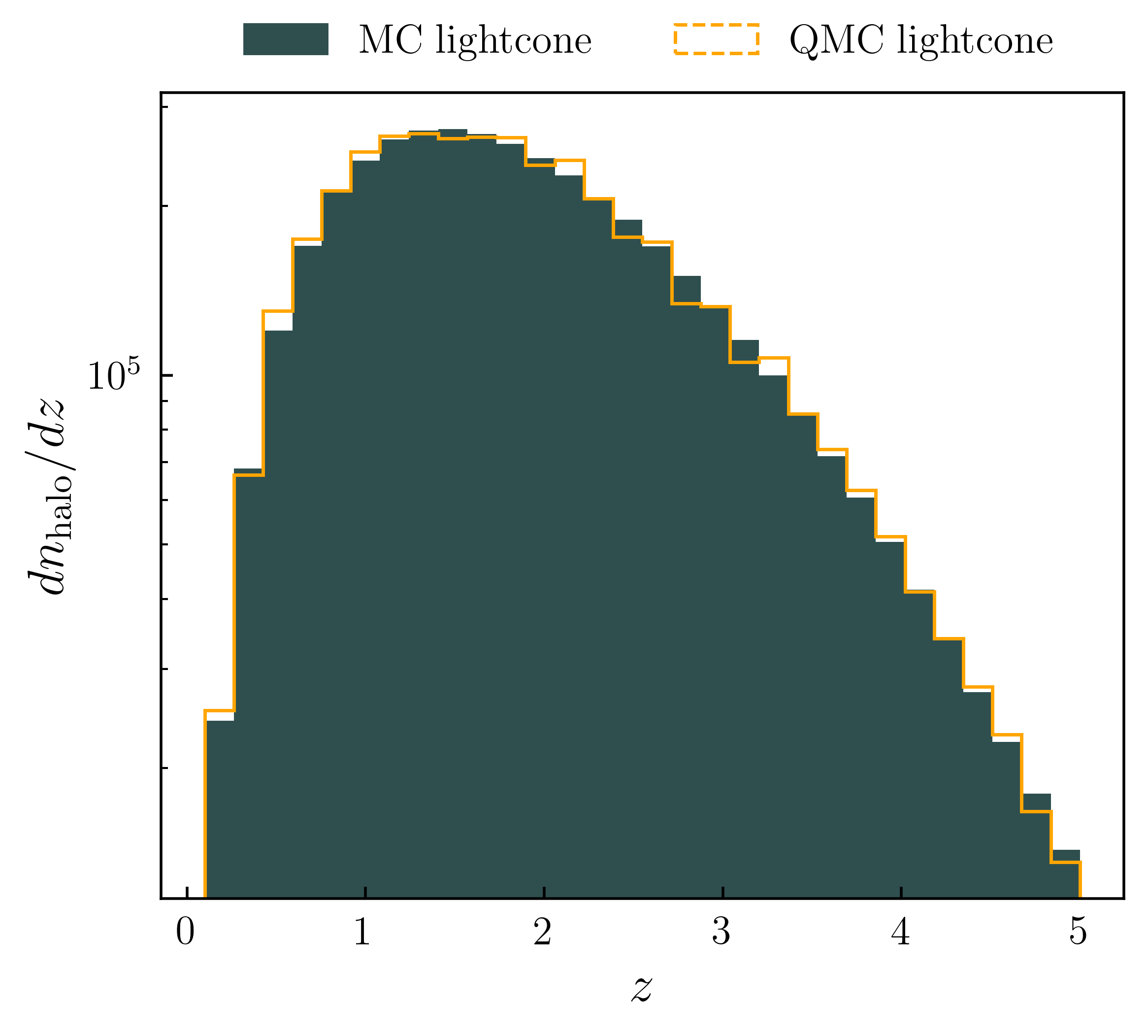}
\caption{\label{fig:hmf_mc_vs_weighted} 
    Example halo lightcones, presented by the redshift distribution $dn_{\rm halo}/dz$ of the halos, generated by Monte Carlo (MC) vs quasi-Monte Carlo (QMC) techniques in the redshift range $z\in[0,5]$. In the MC method, the entire halo population in the lightcone volume is generated; the QMC lightcone is a fast and lightweight alternative based on a weighted grid. See \S\ref{sec:lightcone_generator} for details.
}
\end{figure}
%%%%%%%%%%%%%%%%%%%%%%%%%%%%%%%

Although the MC-generation technique outlined above is computationally efficient, the sheer number of dark matter halos occupying a large cosmological volume can easily produce populations so large as to overflow available memory resources, particularly for small values of $M_{\rm min}.$ For such calculations, \diffhalos includes a \textit{weighted} version of the lightcone generator that follows a Quasi-Monte Carlo (QMC) approach to generating halos, as we briefly describe here. First, we generate a set of $N_{\rm grid}$ points distributed uniformly in mass $m_{ {\rm halo, i}}$ and redshift $z_{\rm i}$ in the range of the lightcone. To each of these points, we assign a weight $w_{\rm i}$ using the differential HMF, $\nhalo(m_{ {\rm halo, i}}, z_{\rm i})$, calculated as the derivative of the cumulative mass function with respect to halo mass, using {\tt jax.grad}. These weights determine the abundance of halos represented by each point, which can then be used to calculate summary statistics based on the QMC-generated halo population. In Figure~\ref{fig:hmf_mc_vs_weighted}, we show a comparison of the redshift distribution of halos generated by the two methods discussed above. As we can see, the two techniques produce statistically consistent samples, with the QMC generator being faster and orders of magnitude more memory efficient. 

%%%%%%%%%%%%%%%%%%%%%%%%%%%%%%%%%%%%%%%%%%%%%%%%%%

\section{Subhalo generator}
\label{sec:subhalo_generator}

Having populated the lightcone with host halos, the next step is to add the population of subhalos to it. In this section, we provide the description of these calculations in \diffhalos. As with the host halo generator in \S\ref{sec:lightcone_generator}, in what follows we introduce a Monte Carlo (MC) and a Quasi-Monte Carlo (QMC) version of the subhalo generator.

To produce lightcones that include subhalos, we use a parameterized model for the \textit{conditional cumulative subhalo mass function} (CCSHMF). Conditioned on $\mhalo$, this model predicts the expected cumulative abundance of subhalos \textit{per individual host} in the lightcone as a function of $\mu \equiv \Msub / \Mhost$, the ratio between the subhalo mass $\Msub$ to the mass $\Mhost$ of its host. We denote this function as $\nsubcuml (>\mu \vert \Mhost)$, and its differential form $\nsub (\mu \vert \Mhost)$ is computed using \texttt{jax.grad} on the cumulative function. 

%%%%%%%%%%%%%%%%%%%%%%%%%%%%%%%
\begin{figure}
\centering
\includegraphics[width=\columnwidth]{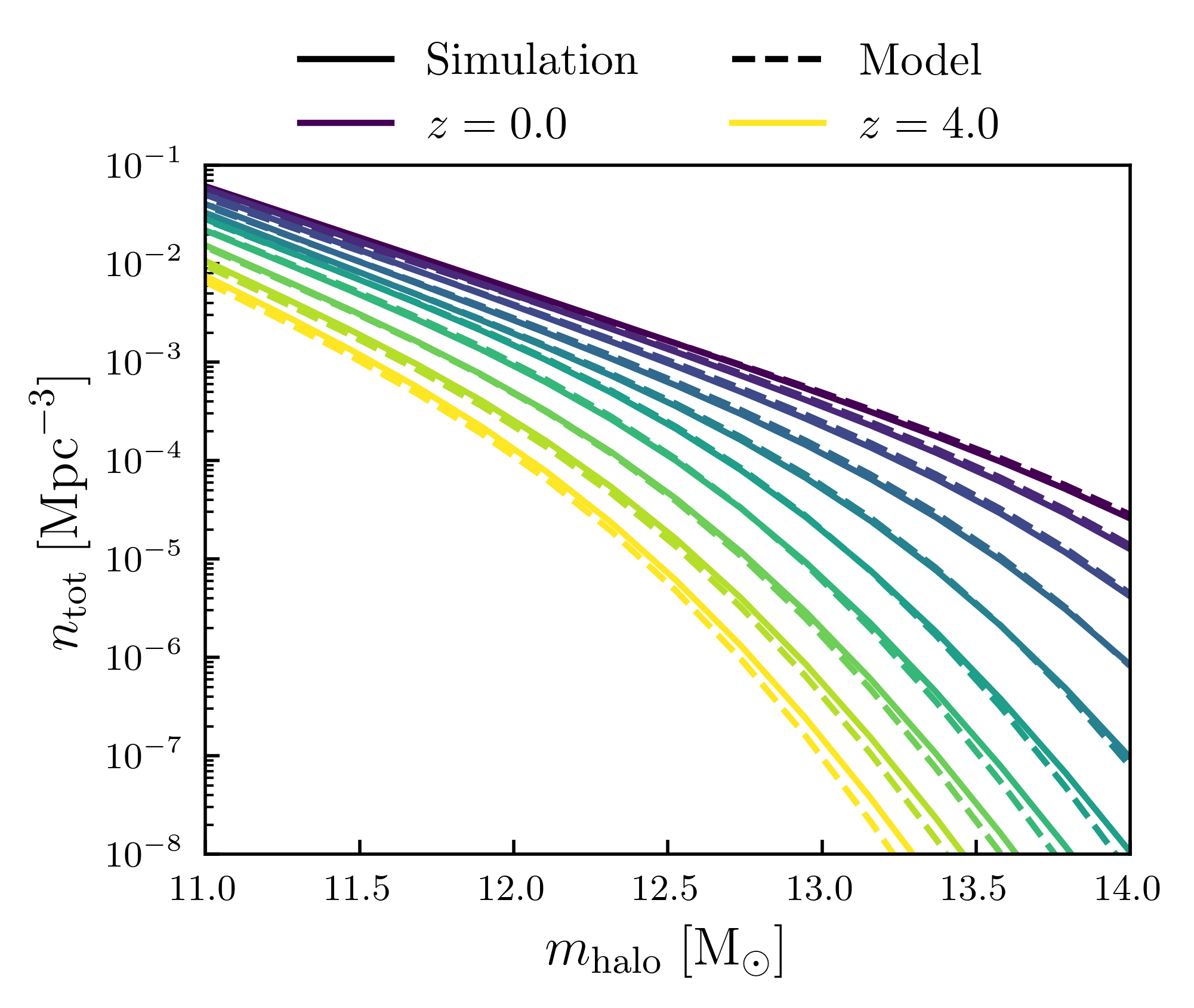}
\caption{\label{fig:CSHMF_model_vs_sim} Comparison of the total mass function of host halos and subhalos between N-body simulations (solid) and the \diffhalos approximation (dashed). Different colored curves show comparisons at different redshifts in the range $0 < z < 4$.}
\end{figure}
%%%%%%%%%%%%%%%%%%%%%%%%%%%%%%%

In \diffhalos, we parametrize the CCSHMF $\nsubcuml (>\mu \vert \Mhost)$ using the same sig-slope kernel we used for the HMF model discussed in \S\ref{sec:HMF}. We tuned this kernel according to the fitting functions developed in \citet{jiang_vdb_2016}, who calibrated their model to match the Bolshoi simulations \citep{klypin_etal11} obtained using the Rockstar \citep{behroozi_etal13_rockstar, behroozi_etal13_consistent_trees} halo finder, over a dynamic range of mass $\sim 10^{11}-10^{15} \; \msun $ and redshift $0-5$. As shown in that paper, these fitting functions provide an accurate description of simulated subhalo mass functions. Additionally, it was shown that the unevolved subhalo mass function has a very weak dependence on redshift. Thus in what follows, we neglect the weak redshift dependence in the CCSHMF, and model the subhalo mass function to vary only according to the mass of the host. To calibrate our model we used the high-resolution, large N-body cosmological Discovery simulations \citep{Beltz-Mohrmann_2025_discovery}, which were carried out with the Hardware/Hybrid Accelerated Cosmology Code\footnote{\url{https://cosmology.alcf.anl.gov/}} (HACC). In Figure~\ref{fig:CSHMF_model_vs_sim} we present the results from our calibration, where we measure and fit our model to the \textit{total mass function} that includes both host halos and subhalos. Our optimized model fits the simulation data reasonably well over the host halo mass range $10 < \log_{10} (\Mhost/\msun) < 14$, and for redshifts between $0 < z < 4$. Further improvements to the calibration could be achieved in concert with a more expansive suite of cosmological simulations (see \S\ref{sec:discussion} for further discussion).

\begin{figure}
\centering
\includegraphics[width=\columnwidth]{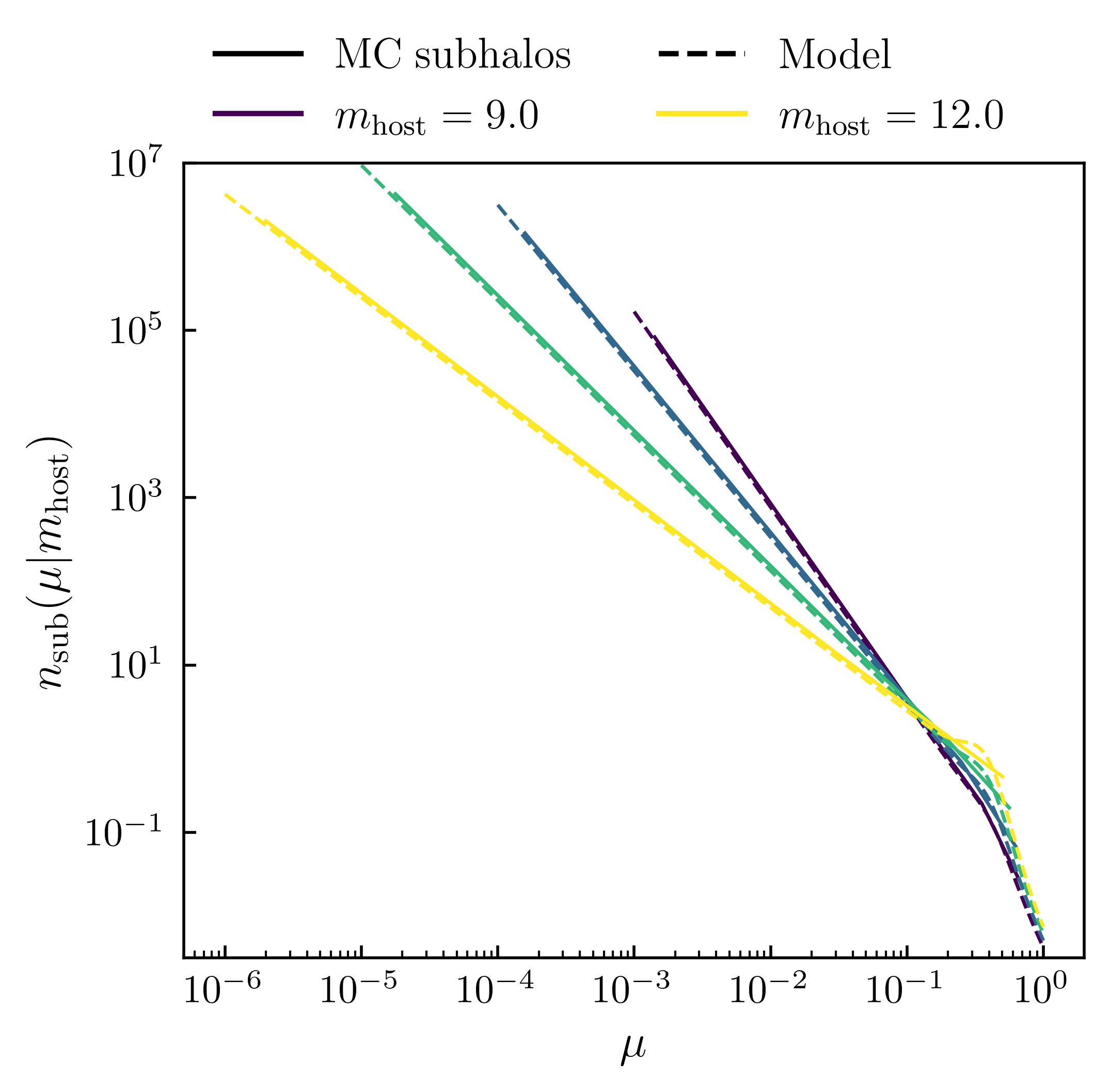}
\caption{\label{fig:CSHMF_model_vs_MCgen} Comparison between the Monte Carlo subhalo generator (solid) and the analytical model prediction (dashed) for subhalos in hosts of different masses, as indicated by the different color lines. The minimum subhalo mass in this case is set to $\msub = 6$.}
\end{figure}

To generate an MC realization of subhalos at redshift $z$, we first generate a host halo population at the same redshift, as described in \S\ref{sec:lightcone_generator}, and then we add subhalos by inverse transform sampling $\nsubcuml(>\mu \vert \Mhost)$, conditioned on the host halo population previously generated. In practice, the process of inverse transformation sampling from the CCSHMF is similar to what was described for sampling from the HMF in \S\ref{sec:lightcone_generator} for hosts. Figure~\ref{fig:CSHMF_model_vs_MCgen} shows that for a MC-generated sample of $10^5$ subhalos, the mass function of the population agrees closely with the underlying analytical model prediction.

%%%%%%%%%%%%%%%%%%%%%%%%%%%%%%%
%%%%%%%%%%%%%%%%%%%%%%%%%%%%%%%

\begin{figure}
\centering
\includegraphics[width=\columnwidth]{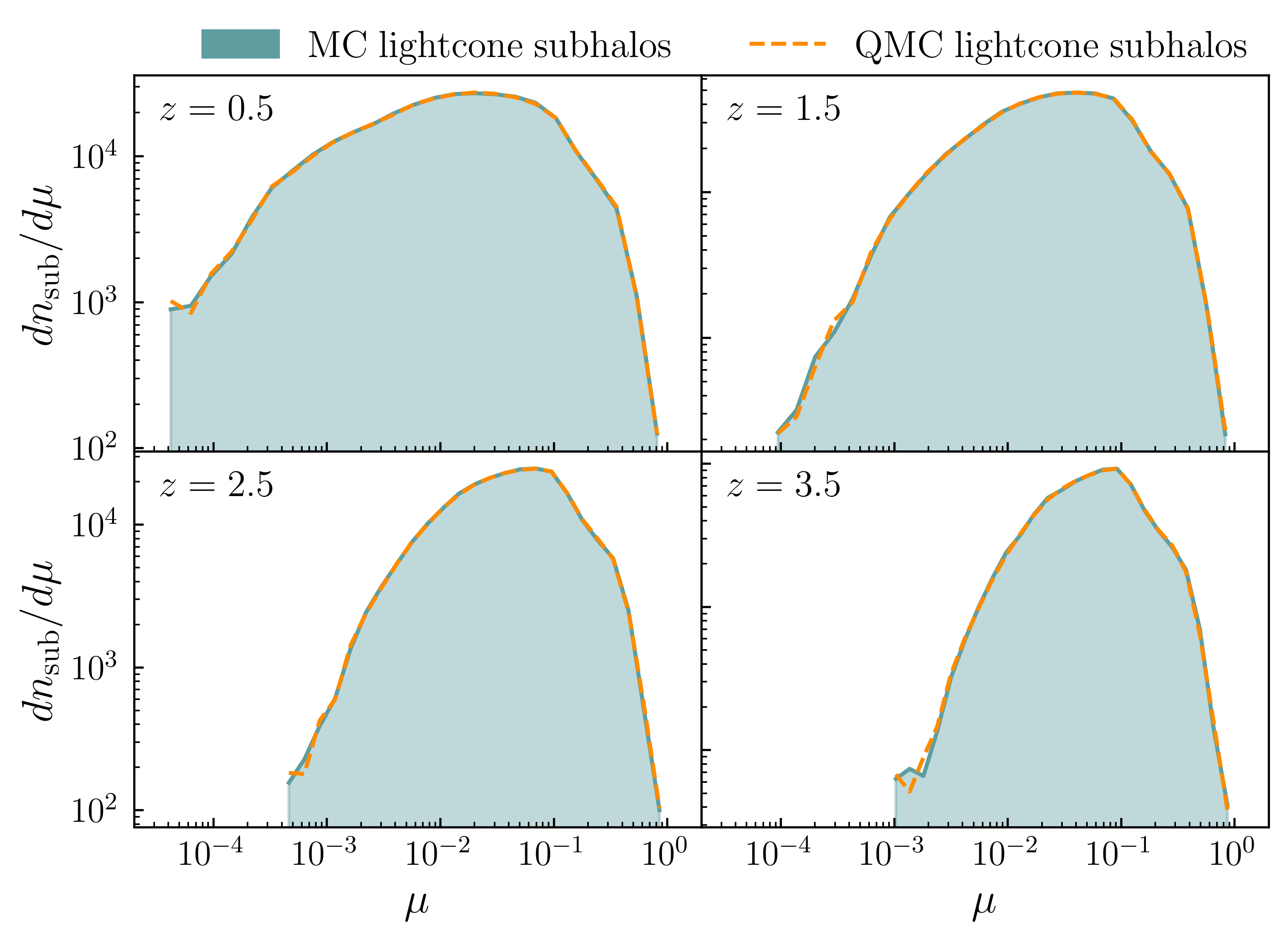}
\caption{\label{fig:lc_subhalo_mc_vs_qmc} Comparison between the Monte Carlo (MC) and quasi-Monte Carlo (QMC) subhalo generator, presented as the distribution of subhalos in $\mu$, for lightcone realizations at different redshifts in each panel.}
\end{figure}

As discussed in \S\ref{sec:lightcone_generator}, MC-generating subhalo populations can easily lead to enormous numbers of objects in typical cosmological calculations. \diffhalos has a QMC method for subhalo generation to address this issue. Briefly, for every host halo of mass $\mhost$, we first generate $N_{\rm grid}$ points $\mu_{\rm i}$ (typically $N_{\rm grid}=5-10$). To each point, we assign a weight $w_{\rm i}\propto\nsub(\mu_{\rm i} \vert m_{\rm host,i}),$ the differential conditional subhalo mass function. In Figure~\ref{fig:lc_subhalo_mc_vs_qmc}, we plot the abundance of subhalos as a function of $\mu$ in lightcones produced using either the MC or the QMC method, showing different redshifts in different panels. Note that this plot shows the total abundance of subhalos in the lightcone, instead of the abundance at a single host halo mass. The QMC and MC lightcones converge for large samples of halos, but the QMC method is faster and can be orders of magnitude smaller in memory footprint.

%%%%%%%%%%%%%%%%%%%%%%%%%%%%%%%
%%%%%%%%%%%%%%%%%%%%%%%%%%%%%%%

%%%%%%%%%%%%%%%%%%%%%%%%%%%%%%%%%%%%%%%%%%%%%%%%%%

\section{DiffmahNet}
\label{sec:diffmahnet}

%%%%%%%%%%%%%%%%%%%%%%%%%%%%%%%
\begin{figure*}
\centering
\includegraphics[]{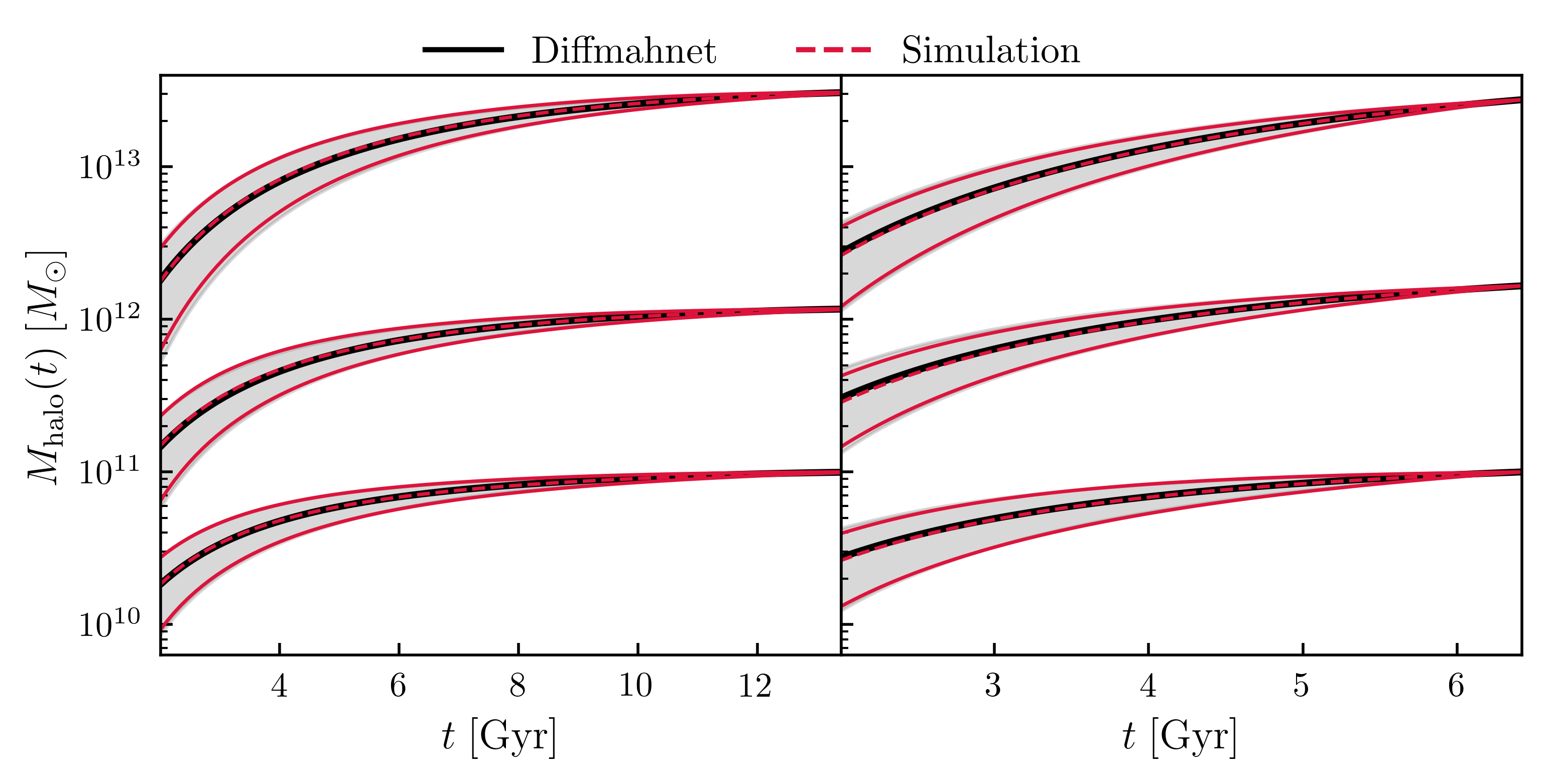}
\caption{\label{fig:diffmahnet_vs_sim} Comparison of the mass assembly histories (MAHs) of halos in the SMDPL N-body simulation vs. \diffmahnet. Each set of curves corresponds to the MAHs of a population of halos identified at time $\tobs$ to have mass $\mobs=\log_{10}M_{\rm halo}(\tobs)/\msun.$ The left panel shows halos identified at $\tobs=13.8\; {\rm Gyr}$, and the right panel halos identified at $\tobs=6.42\; {\rm Gyr}.$ Each panel has three sets of curves corresponding to $\mobs=11$, 12.2, and 13.4.  Each red dashed curve shows the average halo mass prior to attaining $\mobs$ for host halos in the SMDPL N-body simulation; each red set of solid curves shows the variance in $\mhalo$ amongst the SMDPL population. \diffmahnet is a normalizing flow trained to approximate $P(\thetaMAH\vert\mobs,\tobs),$ where $\thetaMAH$ are the \diffmah parameters that determine $\Mhalo(t).$ The black solid curves and gray bands show the mean and variance of the mass of the halo population generated by \diffmahnet. The figure shows that \diffmahnet faithfully reproduces both average halo growth over time, as well as the diversity of halo assembly histories.}
\end{figure*}
%%%%%%%%%%%%%%%%%%%%%%%%%%%%%%%

Having a lightcone that includes populations of host halos and their resident subhalos, we finally assign a mass assembly history (MAH) to each (sub)halo. To achieve this, we use \diffmahnet\footnote{\url{https://diffmahnet.readthedocs.io/en/latest/}}, a population-level model for MAH of halos, conditioned on the mass of the halo $M_{\rm obs} \equiv \Mhalo(t_{\rm obs})$ at the time of observation $t_{\rm obs}$. \diffmahnet is a Python library that produces samples of MAHs for halos, all of which satisfy the above condition that at $t_{\rm obs}$ their mass is $M_{\rm obs}$. In particular, \diffmahnet uses normalizing flows that are trained to emulate \diffmah MAHs for individual dark matter halos. To train the normalizing flows, we used publicly available merger trees in the SMDPL N-body simulation \citep{klypin_etal11}, with halos, subhalos and merger trees identified with Rockstar and ConsistentTrees \citep{behroozi_etal13_rockstar, behroozi_etal13_consistent_trees, rodriguez_puebla_etal16}, as discussed in \S\ref{sec:sims}, which provides a link to the publicly available data. 

In Figure~\ref{fig:diffmahnet_vs_sim}, we compare the predictions from \diffmahnet for three MAHs conditioned on different values of $M_{\rm obs}$, for two cases where $t_{\rm obs} = 13.8 \; {\rm Gyr}$ and $t_{\rm obs} = 6.42 \; {\rm Gyr}$, with data from simulations the model was trained on. The three dashed red curves show the mean MAH for halo populations with different present-day mass in the simulations; the set of solid red curves surrounding each dashed curve shows the variance in the MAH. Each \diffmahnet prediction shown with the corresponding solid black curve, surrounded by the gray band, displays close agreement with the target data from the simulation. Figure~\ref{fig:diffmahnet_vs_sim} demonstrates that \diffmahnet accurately captures both the average halo growth across cosmic time, as well as the diversity of MAHs presented by halos in SMDPL. We note that \diffmahnet has only been trained for the single cosmology of the SMDPL simulation; in future work based on a cosmological suite of N-body merger trees, we aim to include cosmology-dependence into \diffmahnet (see \S\ref{sec:discussion} for further discussion).

%%%%%%%%%%%%%%%%%%%%%%%%%%%%%%%%%%%%%%%%%%%%%%%%%%

\section{Discussion and Future Work}
\label{sec:discussion}

In this paper, we have introduced \diffhalos, a tool to generate cosmological lightcones of dark matter halos, subhalos, and their mass assembly histories. In developing this library, our own primary motivation was to support ongoing work building and calibrating models of the galaxy--halo connection. In \citet{Alarcon_2025_diffstarpop}, the \DstarPop model of galaxy star formation history (SFH) was shown to be flexible enough to approximate populations of SFH taken from a wide range of simulations, and computationally efficient enough for applications of Bayesian inference such as MCMC. We are currently using \DstarPop as the SFH ingredient in Diffsky, a forward modeling pipeline for predicting galaxy spectral energy distributions (SEDs), and optimizing the model predictions according to observational measurements of galaxy SEDs and photometry. Using \diffhalos-generated lightcones instead of N-body-simulated halos dramatically reduces the size and runtime of the computations needed to optimize the Diffsky model parameters. Semi-analytic models (SAMs) of galaxy formation have used similar techniques to speed up the optimization of SAM parameters. For example, downsampled and upweighted N-body merger trees were used to optimize the parameters of the Munich SAM presented in \citet{henriques_etal_2015}; in \citet{robertson_benson_2025} the Galacticus SAM parameters were optimized by evaluating the model at a small set of target halo masses; in \citet{Pandya_etal2026_sapphire} sample halo merger trees used to optimize the model were selected with a latin hypercube design. For applications optimizing models of the galaxy--halo connection, \diffhalos provides a programmatic way to generate small samples of halos that span the necessary range of halo mass, redshift, and assembly history, and that can be abundance-weighted when making predictions for observed quantities.

The cosmology dependence of \diffhalos-generated lightcones is captured via the halo mass function emulator we have built specifically for the proof-of-concept applications in this paper. In future applications, incorporating one of the numerous state-of-the-art mass function emulators \citep[e.g.,][]{Heitmann_etal2013_Coyote,Angulo_etal2020_baccoemu,Knabenhans_etal2020_EuclidEmulator2,Moran_etal2022_MiraTitan,brieden_etal25_web_halo_model,lovell_etal26}, or fitting functions  \citep{Benson_etal_2026Unified_HMF,zheng_etal26_hmf_all_mass}, into our pipeline can significantly increase the accuracy and flexibility of \diffhalos. Calibrating the cosmology-dependence of the subhalo mass function and the halo MAHs will also be essential to enabling precision studies of cosmology. Such calibrations will require an extensive suite of high-resolution N-body simulations with substructure merger trees, utilized in concert with addressing formidable numerical challenges of subhalo finding \citep[see, e.g.,][]{vdb_etal18a,vdb_etal18b}, and possibly adopting halo-finders that improve the separation of bound from unbound particles by using dynamics-based halo definitions \citep{diemer_22_dynamical_halos,garcia_etal23_dynamical_halos}, or by leveraging the boosted gravitational potential \citep{Richardson_etal26_strawberry}.
Significant progress in this direction has been made recently that can support our efforts to calibrate the subhalo mass function in \diffhalos. For example, {\tt SYMFIND} \citep{mansfield_etal24_symfind}, a particle-tracking-based subhalo finder that is capable of tracking subhalos with improved robustness than commonly used algorithms such as Rockstar, can play a key role in calibrating the cosmology dependence of the subhalo mass function in our model. The {\tt SYMFIND} halo finder was developed in part using {\tt Symphony} \citep{nadler_etal23_symphony1}, a suite of zoom-in cosmological simulations that provide the necessary data products that we could utilize to study how the conditional subhalo mass function depends on host halo mass and assembly history. In closely related work, \cite{Bera_Diemer_2026} leveraged the Erebos \citep{diemer_sparta_erebos_2020}, IllustrisTNG \citep{Springel_2018_IllustrisTNG,Pillepich_2018_IllustrisTNG,Nelson_2018_IllustrisTNG,Naiman_2018_IllustrisTNG,Marinacci_2018_IllustrisTNG}, and Thesan \citep{Kannan_2022_Thesan,Smith_2022_Thesan,Garaldi_2022_Thesan} simulations to make measurements of the mass accretion rate of halos in dark matter-only and hydrodynamical simulations in different cosmologies, and over a wide range of redshift between $0 < z < 14$. Using their results, we could calibrate the cosmology-dependence of the predicted MAHs in our generated lightcones in future applications. While further improvements upon the calibration of \diffhalos will be needed for precision cosmology, the level of accuracy achieved here may already be sufficient to support the development and calibration of models of galaxy photometry: present-day uncertainty in stellar population synthesis ingredients such as the assumed dust model \citep{hahn_melchior_2025_dust_biases} is likely the dominant systematic in forward modeling pipelines.

In closely related work, the semi-analytic model {\tt SatGen}  \citep{satgen_Jiang_etal_2021} has the capability to generate satellite populations for host halos of a given mass, redshift, and MAH. {\tt SatGen} uses the EPS framework to generate merger trees, and takes into account baryonic effects on both host halos and their satellites. In future extensions of \diffhalos, combining our framework together with orbital evolution models such as those in {\tt SatGen} would enhance the predictions of our model to include satellite merging and radial distributions in a manner that accounts for baryonic effects seen in hydrodynamical simulations. In this direction, we can further increase the realism of \diffhalos-generated subhalo populations by incorporating into our model how baryonic effects influence subhalo disruption and merging \citep[e.g.,][]{nadler_etal18}, and how different dark matter models alter the shape of the subhalo mass function \citep[e.g.,][]{nadler_etal21}.

Finally, we highlight a limitation of the current Diffsky-\diffhalos framework: the sky lightcone coordinates of our generated halos only includes redshift, but not \{ra, dec\}. To address this in future work, we aim to expand \diffhalos to enable the capability of jointly predicting the cosmic density field. For instance, we can achieve this by incorporating \diffhalos into differentiable simulators \citep[e.g.,][]{Li_etal22_pmwd,horowitz_etal25_diffhydro}, or by coupling \diffhalos with other techniques for generating the density field \citep[e.g.,][]{mustafa_etal19_cosmogan,kodi_ramanah_etal20}. With this application squarely in mind, we have intentionally formulated \diffhalos to facilitate integration into existing pipelines. For example, DISCO-DJ \citep{hahn_etal24_discodj1,list_etal2025_discodj2} generates both the density field and a population of host halos; in this case, \diffhalos could augment the host halo population created by DISCO-DJ with subhalos and MAHs, simplifying the task of integrating the two approaches to the problem. Including the capability to predict clustering statistics into \diffhalos will increase the constraining power of our model on cosmology and the galaxy--halo connection, and  provide a powerful framework for cosmological inference by the present-- and next--generation surveys of the large-scale structure.

%%%%%%%%%%%%%%%%%%%%%%%%%%%%%%%%%%%%%%%%%%%%%%%%%%

\section{Conclusions}
\label{sec:conclusion}

In this paper, we have introduced \diffhalos: a model for generating cosmological lightcones of dark matter halos, subhalos, and their mass assembly histories (MAHs). Our model is written in \jax for automatic differentiation, which enables gradient-based applications of \diffhalos, such as gradient descent and Hamiltonian Monte Carlo techniques. Our main results are the following:
\begin{enumerate}
    \item\diffhalos can generate Monte Carlo (MC) realizations of populations of host halos in a cosmological lightcone, with mass and redshift distributions that are in close agreement with theoretical expectations. We have additionally introduced a quasi-Monte Carlo (QMC) technique that reduces the memory footprint of the halo lightcones by orders of magnitude.

    \item \diffhalos also generates subhalos that populate the halos in the lightcone, achieved through a model of the conditional subhalo mass function.

    \item Each host halo and subhalo in a \diffhalos-generated lightcone is bestowed with a MAH by \diffmahnet, a normalizing flow that captures the probability distribution of MAHs conditioned on redshift and halo mass at the time of observation. 
    
\end{enumerate}

This paper presents a proof-of-concept of the \diffhalos framework, and supplies a publicly available tool, \url{https://github.com/ArgonneCPAC/diffhalos}, that can be used to support the development and calibration of models of the galaxy--halo connection.

%%%%%%%%%%%%%%%%%%%%%%%%%%%%%%%%%%%%%%%%%%%%%%%%%%

\section*{Acknowledgements}

We thank Ankita Bera and Benedikt Diemer for comments on an early draft of this manuscript.

We thank the developers of {\tt NumPy} \citep{numpy_ndarray}, {\tt SciPy} \citep{scipy}, Jupyter \citep{jupyter}, IPython \citep{ipython}, scikit-learn \citep{scikit_learn}, \jax \citep{jax2018github}, and Matplotlib \citep{matplotlib} for their extremely useful free software. While writing this paper we made extensive use of the Astrophysics Data Service (ADS) and {\tt arXiv} preprint repository. The authors gratefully acknowledge the Gauss Centre for Supercomputing e.V. (www.gauss-centre.eu) and the Partnership for Advanced Supercomputing in Europe (PRACE, www.prace-ri.eu) for funding the MultiDark simulation project by providing computing time on the GCS Supercomputer SuperMUC at Leibniz Supercomputing Centre (LRZ, www.lrz.de). The Bolshoi simulations have been performed within the Bolshoi project of the University of California High-Performance AstroComputing Center (UC-HiPACC) and were run at the NASA Ames Research Center.

Work done at Argonne National Laboratory was supported under the DOE contract DE-AC02-06CH11357. This work was supported in part by the OpenUniverse effort, which is funded by NASA under JPL Contract Task 70-711320, ``Maximizing Science Exploitation of Simulated Cosmological Survey Data Across Surveys". We gratefully acknowledge use of the Bebop cluster in the Laboratory Computing Resource Center at Argonne National Laboratory.

\bibliographystyle{aasjournal}
\bibliography{bibliography}

\end{document}